\documentstyle[fleqn,12pt]{article}

\begin{document}
\def\uslash{{U\mskip -12mu/}}
\begin{quote}
\center{\Large{\bf The Role of Zero-Modes in
the Canonical Quantization of Heavy-Fermion QED
in Light-Cone Coordinates}}
\vskip4ex
\center{{\sf Robert W. Brown}, {\sf Jin Woo Jun${\,\!}^\dagger$},
{\sf Shmaryu M. Shvartsman}} and {\sf Cyrus C. Taylor}
\vskip2ex
\center{{\sl Department of Physics, Case Western Reserve University, \\
Cleveland, Ohio 44106}}
\vskip5ex
\end{quote}
\bigskip\bigskip
\begin{abstract}

	Four-dimensional heavy-fermion QED is studied in
light-cone coordinates with (anti-)periodic field boundary conditions.
We carry out a consistent light-cone  canonical quantization of
this model using the Dirac algorithm for a system with first- and
second-class
constraints. To examine the role of the zero modes,
we consider the quantization procedure in
{the }zero-mode
{and the non-zero-mode} sectors separately. In both sectors we
obtain the physical
variables and their canonical commutation relations. The physical
Hamiltonian is constructed via a step-by-step exclusion of the
{unphysical degrees of freedom.}
An example using this Hamiltonian in which the zero
modes play a role is the verification of the correct Coulomb
potential between two heavy fermions.
\end{abstract}
\vspace{1.75in}
$\dagger$ Permanent address: Department of Physics, Inje University,\\
Kimhae, 621-749, Republic of Korea
\section{Introduction}

\hspace{3em}There has been recent progress [1-4]
towards a framework for describing decay
processes involving heavy quarks.  Its basis has been
the study of QCD in which some of the quark masses are taken to
infinity.  In this limit, the quark spin degrees of freedom decouple
from each other and the
couplings of the heavy quarks to the gluon degrees of freedom become
mass independent and are described by a Wilson line.

The remarkable consequence of this independence is that all matrix
elements of vector and axial vector currents bilinear in either or both
of a pair of heavy quarks taken between either vector or pseudoscalar
mesons in initial and final states at arbitrary momentum transfer are
determined in terms of a single normalized function.  The question
arises as to how one calculates this function.

One of the more promising approaches to problems in
QCD is light-cone quantization \cite{brev1,brev2}.
Light-cone quantization has turned
out to be a useful tool
for the perturbative treatment of field theories \cite{Brodsky,Brodsky1}.
In its extension
[9-13] to the
nonperturbative domain, we have come to
realize that  careful attention must be paid to
the non-trivial vacuum structure of light-cone quantum field theory.
For instance the
light-cone vacuum in
the massless Schwinger model can be understood only
by a careful study of the zero-modes
of the constraints imposed by the light-cone frame
[14-16].  Indeed it has
been conjectured that the dynamics of zero modes in QCD in
light-cone quantization provide the mechanism for confinement
\cite{brev1,brev2}.

	In the present paper we consider a four-dimensional
heavy-fermion QED, as an initial step towards the study of heavy-quark
QCD in light-cone quantization.
We carry out a consistent canonical quantization of this
gauge invariant model in a light-cone domain restricted in its
``spatial'' directions. It is well known that in such a restricted
region one has problems with zero modes
[17,15,16].
The canonical quantization of the massless Schwinger model on a
light-like hyperplane, taking into account zero-mode contributions,
was considered in \cite{Heinzl,Heinzl2}. Neglecting such
contributions, Tang {\it et al.} \cite{Tang} carried out the
discretized light-cone quantization of four-dimensional QED and
Mustaki \cite{Mustaki} developed the canonical quantization of
two-dimensional QED on the null plane.  We note that these questions
have also been investigated by first quantizing and then taking
the heavy-quark limit, in two-dimensional QCD
\cite{burkardt}.

	In studying the quantization of gauge field theories
one is confronted by first-class constraints and their
corresponding gauge conditions.
A consistent canonical quantization formalism for  such problems was
proposed long ago
by Dirac \cite{Dirac} and Bergmann \cite{Bergmann}, and its
generalization to fermionic (Grassmann-odd) constraints by Casalbuoni
\cite{Casalbuoni}. (There are in fact books on this subject
\cite{Sundermayer,Gitman}. We will not discuss here other approaches
involving path integration, BRST or BFV methods.)
In the Dirac approach some problems arise. They involve the
determination of the dynamical (physical) variables and the construction
of the physical Hamiltonian. One needs to prove, also, that in
the physical sector the $S$-matrix does not depend on the form of
the gauge conditions.

	The specific gauge theory we address in this paper is in terms
of light-cone variables where, as we shall see, the quantization comes
with some important constraints involving zero-frequency variables which
require special attention.  To examine the role of the zero modes
explicitly we consider the
quantization procedure in the separated zeroless-mode and zero-mode
sectors. In such sectors we choose special gauge conditions and obtain
the physical variables and their canonical commutation relations. The
physical Hamiltonian is constructed by systematic elimination of the
nondynamical variables.

	In Sections 2-5 the canonical quantization of heavy
fermion QED is addressed by way of the Dirac-Bergmann
algorithm. In Section 6
we calculate the interaction potential between a heavy fermion and a heavy
antifermion using old-fashioned
perturbation theory. Conclusions and a brief discussion follow in Section 7.

\vskip2ex

\setcounter{equation}{0}
\section{Constrained Dynamics of Heavy Fermion QED}
\vskip2ex

\hspace{3em}We begin with the Lagrangian of heavy-fermion QED
obtained \cite{Isgur} by a generalized
Foldy-Wouthuysen transformation
that removes the terms mixing fermion and anti-fermion fields in
the action.  We have
\begin{equation}
{\cal L}=i\overline{\Psi}\uslash {U}^{\mu}D_{\mu}\Psi-M\overline{\Psi}\Psi-
\frac{1}{4}F_{\mu\nu}F^{\mu\nu},
\label{cd1}
\end{equation}
where the Minkowski metric is diag$g_{\mu \nu}=(1,-1,-1,-1)$,
\begin{equation}
D_{\mu}=\partial_{\mu}+ieA_{\mu}
\label{cda}
\end{equation}
is the covariant derivative,
\begin{equation}
\uslash {U}^{\mu}=\gamma\cdot U{U}^{\mu},
\label{cdb}
\end{equation}
and $U^{\mu}$ is the given
4-velocity of the heavy fermion satisfying the condition $U^{2}=1$.
The field strength tensor is
\begin{equation}
F_{\mu\nu}=\partial_{\mu}A_{\nu}-\partial_{\nu}A_{\mu},
\label{cd1a}
\end{equation}
and we use the system of units where $\hbar=c=1$.
The heavy fermion limit means that $MU^{\mu}$ is
greater than any other momentum in the problem under consideration.

	The light-cone coordinates in four-dimension are
$x^{\mu}=(x^{+},x^{-},x^{j})$, $j=1,2$, where
\begin{equation}
x^{\pm}=\frac{1}{\sqrt{2}}(x^{0}\pm x^{3})
\label{cdbc}
\end{equation}
The variable $x^{+}$ plays the role of the ``time'' variable.
In terms of such coordinates the Lagrangian ${\cal L}$  becomes
\begin{eqnarray}
{\cal L} &=& i\overline{\Psi}\uslash
 ({U}_{+}\partial_{-}+{U}_{-}\partial_{+}-
 {U}_{j}\partial_{j})\Psi-M\overline{\Psi}\Psi
+\frac{1}{2}F^{2}_{+-}+F_{+j}F_{-j}-\frac{1}{4}F_{jk}F_{jk}\nonumber \\
& &-e\overline{\Psi}\uslash ({U}_{+}A_{-}+
{U}_{-}A_{+}-{U}_{j}A_{j})\Psi,
\label{cd0}
\end{eqnarray}
where
\begin{eqnarray}
&F_{jk}=\partial_{j}A_{k}-\partial_{k}A_{j},\nonumber \\
&F_{+-}=\partial_{+}A_{-}-\partial_{-}A_{+},\nonumber \\
&F_{\pm j}=\partial_{\pm}A_{j}-\partial_{j}A_{\pm}, \nonumber \\
&A_{\mp}=\frac{1}{\sqrt{2}}(A^{0}\pm A^{3}),\nonumber \\
&{U}_{\mp}=\frac{1}{\sqrt{2}}({U}^{0}\pm {U}^{3}),\nonumber \\
&\partial_{\pm}=\partial /\partial x^{\pm}
\label{cd0a}
\end{eqnarray}

	The fermion part of the Lagrangian (\ref{cd1}) differs from ordinary
QED by the change
$\gamma^{\mu}\rightarrow \uslash {U}^{\mu}$. Let us consider the equations
of motion for the fermion field $\Psi$. They are
\begin{equation}
i\uslash {U}^{\mu}D_{\mu}\Psi=M\Psi
\label{cd01}
\end{equation}
As in ordinary QED, let us introduce, in light-cone coordinates, two
``chiral-like" fermion fields [26-29]
\begin{eqnarray}
\Psi_{(\pm)}&=&P_{(\pm)}\Psi,\nonumber \\
P_{(\pm)}&=&\frac{1}{2}(1\pm \gamma_{0}\gamma _{3})
\label{cd02}
\end{eqnarray}
Since $U^{2}=1$ the equations (\ref{cd01}) may be rewritten in the form
\begin{equation}
iU^{\mu}D_{\mu}\Psi_{(\pm)}=M(\sqrt{2}U_{\pm}\gamma_{0}\Psi_{(\mp)}-
U_{j}\gamma_{j}\Psi_{(\pm)})
\label{cd03}
\end{equation}
{}From these equations it follows that both fields $\Psi_{(\pm)}$ are
dynamical fields because both of them contain the derivative with
respect to the ``time'' coordinate
$x^{+}$. This distinguishes heavy-fermion QED from
ordinary QED where the fields $\Psi_{(\pm)}$ are not independent
[26-29] in light-cone quantization:
The field $\Psi_{(+)}$ can be related to the dynamical field
$\Psi_{(-)}$ in ordinary QED.  (See, however, the discussion of zero-modes
in \cite{mccartor2}.)

	The Lagrangian (\ref{cd1}) is gauge invariant.
This means that the classical theory contains ``first-class'' constraints
and we need a quantization prescription for  systems with constraints
such as that provided by Dirac \cite{Dirac}.

	According to this procedure ``primary constraints'' arise when
one cannot relate a velocity to its corresponding canonical momentum.
The consistency condition of a primary constraint, which means that the
constraint must be conserved in time, is used either as a condition which
defines a Lagrange multiplier function or as a ``secondary'' constraint.
There could  even be ``tertiary constraints'', and so on.
This procedure terminates when no new constraints appear.

	All constraints are divided further into two groups. Constraints
whose Poisson brackets with all other constraints vanish on the
constraint surface are
called  ``first-class'' constraints. Otherwise, they are called
``second-class'' constraints.
If a constraint is first class, a subsidiary
condition (gauge condition) must be imposed in order to determine the
corresponding Lagrange multiplier function. When all constraints finally
become second class, we can invert the matrix of Poisson
brackets between the constraints and replace the
Poisson bracket by the Dirac bracket. The quantization procedure
consists of replacement of the Dirac bracket by a commutator for bosons or
an anticommutator for fermions.

	Let us consider the quantization of  the theory in the
restricted region \\$-L\leq  x^{-}\leq  L$,
$-L\leq x^{j}\leq L$ and impose periodic boundary condition for
the bosonic variables and antiperiodic boundary condition for the
fermionic ones. In such a
restricted region, the fact that the classical equations are first order
in the ``time'' $x^{+}$ leads to the well known problem of zero modes. To
separate the zero mode contributions, it is useful
to generalize the orthogonal projection operators found in
 \cite{Heinzl}
\begin{eqnarray}
{\cal P}_{(a)}\ast \phi(\underline{x})&=&\int_{-L}^{L}
{\cal P}_{(a)}(\underline{x},\underline{y})\phi(\underline{y})d\underline{y},
\nonumber\\
{\cal P}_{(a)}&=&({\cal P},{\cal P}_{j},{\cal Q},
{\cal Q}_{j}),\;j=1,2,\nonumber \\
{\cal Q}&=&{\cal I}-{\cal P},{\cal Q}_{j}={\cal I}-{\cal P}_{j},\nonumber \\
{\cal P}(\underline{x},\underline{y})&=&\frac{1}{2L}\delta(x_{\perp}-
y_{\perp}),\nonumber \\
{\cal P}_{j}(\underline{x},\underline{y})&=&\frac{1}{2L}\epsilon_{jk}
\delta(x^{k}-y^{k})\delta(x^{-}-y^{-}),
\label{cd4}
\end{eqnarray}
where $\underline{x}=(x^{-},x^{1},x^{2})=(x^{-},x_{\perp})$, antidiag
$\epsilon_{jk}=(1,1)$, and ${\cal I}$ is a unit operator:\\ ${\cal I}
\ast\phi(\underline{x})=
\phi(\underline{x})$.
Here ${\cal P},{\cal P}_{j}$ are the
projection operators into the zero-mode sectors of $x^{-},x^{j}$,
respectively, and ${\cal Q},{\cal Q}_{j}$ are the
projection operators which eliminate the zero modes in the sectors of
$x^{-},x^{j}$, respectively. We shall speak of $\cal P$ as the
projector onto the $\cal P$-sector, and $\cal Q$ as the projector
onto the $\cal Q$-sector.

In the same way it is possible to define the
product of two projectors
\begin{equation}
{\cal P}_{(a)}{\cal P}_{(b)}\ast\phi(\underline{x})={\cal P}_{(a)}\ast
({\cal P}_{(b)}\ast\phi)(\underline{x})=
{\cal P}_{(b)}{\cal P}_{(a)}\ast\phi(\underline{x})
\label{cd4a}
\end{equation}
Due to this definition, the operator ${\cal P}_{1}{\cal P}_{2}$ is
orthogonal to ${\cal Q}_{1},{\cal Q}_{2}$ and ${\cal Q}_{1}
{\cal Q}_{2}$. One
can verify that the operators ${\cal P}_{1}{\cal P}_{2}$ and
${\cal Q}_{1}+{\cal Q}_{2}-{\cal Q}_{1}{\cal Q}_{2}$ are projection
operators.  Further, ${\cal P}_{1}{\cal P}_{2} +
{\cal Q}_{1}+{\cal Q}_{2}-{\cal Q}_{1}{\cal Q}_{2}=1$, so that the
two subspaces defined by these projection operators span the space.

	We define the canonical momenta $\Pi_{\phi}$  as
\begin{eqnarray}
& &\Pi_{\phi}=\frac{\partial_{r}{\cal L}}{\partial \dot{\phi}},
\nonumber \\
& &\dot{\phi}=\partial_{+}\phi
\label{cd4b}
\end{eqnarray}
Then
\begin{eqnarray}
& &\Pi_{+}(\equiv \Pi_{A_{+}})=0,\nonumber \\
& &\Pi_{-}(\equiv \Pi_{A_{-}})=\dot{A}_{-}-\partial_{-}A_{+},\nonumber\\
& &\Pi_{j}(\equiv \Pi_{A_{j}})=\partial_{-}A_{j}-\partial_{j}A_{-},
\nonumber \\
& &\Pi_{\Psi}=i\overline{\Psi}\uslash {U}_{-},\nonumber \\
& &\Pi_{\overline{\Psi}}=0
\label{cd3}
\end{eqnarray}
The velocity $\dot{A}_{-}$ can be expressed through the momentum
$\Pi_{-}$ and we have five primary constraints ${\bf \Phi^{(1)}}\approx
0$, where
\begin{equation}
{\bf \Phi^{(1)}}=\left\{ \begin{array}{ll}
                 \phi^{(1)}_{+}=\Pi_{+}\\
                 \phi^{(1)}_{j}=\Pi_{j}-\partial_{-}A_{j}+\partial_{j}A_{-}\\
                 \phi^{(1)}_{\Psi}=\Pi_{\Psi}-i\overline{\Psi}\uslash {U}_{-}\\
                 \phi^{(1)}_{\overline{\Psi}}=\Pi_{\overline{\Psi}}
                 \end{array}
         \right.
\label{cd3a}
\end{equation}
The canonical Hamiltonian density on the constraint surface (\ref{cd3a})
is
\begin{eqnarray}
{\cal H}_{c}&=& -\Pi_{\Psi}{U}^{-1}_{-}({U}_{+}\partial_{-}-
{U}_{j}\partial_{j})\Psi-iM\Pi_{\Psi}\uslash{U}^{-1}_{-}
\Psi \nonumber \\
& &-ie\Pi_{\Psi}{U}^{-1}_{-}{U}^{\mu}A_{\mu}\Psi
+\frac{1}{2}\Pi_{-}^{2}-
 (\partial_{-}\Pi_{-}+\partial_{j}\Pi_{j})A_{+}+
\frac{1}{4}F_{jk}F_{jk}
\label{cd2p}
\end{eqnarray}

	Consider the Poisson brackets between the constraints
$\phi^{(1)}_{j}$ from (\ref{cd3a})
\begin{equation}
\left\{\phi^{(1)}_{j}(\underline{x}),\phi^{(1)}_{k}(\underline{y})\right\}=
-2\delta_{jk}\partial_{-}\delta(\underline{x}-\underline{y})
\label{cd3b}
\end{equation}
{}From this expression we see that the constraints $\phi^{(1)}_{j}$
commute  only in the ${\cal P}$-sector. Hence in the ${\cal P}$-sector
the consistency
conditions for the constraints $\phi^{(1)}_{j}$ may lead to secondary
constraints. Thus let us consider the decomposition of the primary
constraints $\phi^{(1)}_{+}$ and $\phi^{(1)}_{j}$ into the ${\cal Q}$ and
${\cal P}$ sector constraints
\begin{eqnarray}
 & & \phi^{(1,{\cal Q})}_{+}={\cal Q}\ast \Pi_{+},
\nonumber \\
 & & \phi^{(1,{\cal Q})}_{j}={\cal Q}\ast
(\Pi_{j}-\partial_{-}A_{j}+\partial_{j}A_{-}),
\label{cd6a}\\
 & & \phi^{(1,{\cal P})}_{+}={\cal P}\ast \Pi_{+},
\nonumber \\
 & & \phi^{(1,{\cal P})}_{j}={\cal P}\ast
(\Pi_{j}+\partial_{j}A_{-})
\label{cd6}
\end{eqnarray}


	Following the Dirac prescription we construct the primary
Hamiltonian density
\begin{eqnarray}
{\cal H}^{(1)}&=&{\cal H}_{c}+\lambda_{\Psi}\phi^{(1)}
_{\Psi}+\lambda_{\overline{\Psi}}\phi^{(1)}_{\overline{\Psi}}+
\lambda^{({\cal Q})}_{+}\phi^{(1,{\cal Q})}_{+}\nonumber \\
& &+\lambda^{({\cal Q})}
_{j}\phi^{(1,{\cal Q})}_{j}+
\lambda^{({\cal P})}_{+}\phi^{(1,{\cal P})}_{+}+\lambda^{({\cal P})}
_{j}\phi^{(1,{\cal P})}_{j},
\label{cd7}
\end{eqnarray}
where $\lambda^{({\cal Q},{\cal P})}_{+},\lambda^{({\cal Q},
{\cal P})}_{j}$
are Grassmann-even
Lagrange multipliers for the constraints (\ref{cd6a}), (\ref{cd6})
and $\lambda_{\Psi},
\lambda_{\overline{\Psi}}$ are Grassmann-odd Lagrange multipliers for the
fermionic constraints $\phi^{(1)}_{\Psi},\phi^{(1)}_{\overline{\Psi}}$,
respectively. The consistency conditions for the constraints
$\phi^{(1)}_{\Psi},\phi^{(1)}_{\overline{\Psi}}$  do not give new
constraints, and hence lead to the determination of their Lagrange
multipliers. We will consider the ${\cal Q}$ and ${\cal P}$ sectors
separately.

\setcounter{equation}{0}
\section{{\cal Q}-sector}
\vskip2ex

\hspace {3em} In this Section we are going to consider the quantization
procedure in the ${\cal Q}$-sector. In this sector the primary
Hamiltonian density is
\begin{equation}
{\cal H}^{(1,{\cal Q})}={\cal H}_{c}+
\lambda^{({\cal Q})}_{+}\phi^{(1,{\cal Q})}_{+}+\lambda^{({\cal Q})}
_{j}\phi^{(1,{\cal Q})}_{j}
\end{equation}
The consistency conditions for the constraints $\phi^{(1,{\cal Q})}_{j}$
determine their Lagrange multipliers, and for the constraint
$\phi^{(1,{\cal Q})}_{+}$ gives a new constraint
\begin{equation}
\phi^{(2,{\cal Q})}_{+}={\cal Q}\ast(\partial_{-}\Pi_{-}+\partial_{j}\Pi_{j}+
ie\Pi_{\Psi} \Psi)
\label{cd8}
\end{equation}
This constraint does not commute with the fermionic constraints. Thus let us
consider a new ``Gaussian law'' constraint which is equivalent to (\ref{cd8})
in the presence of (\ref{cd3a})
\begin{equation}
\phi^{(2,{\cal Q})}_{+}={\cal Q}\ast(\partial_{-}\Pi_{-}
+\partial_{j}\Pi_{j}+ ie\Pi_{\Psi} \Psi+ ie \overline{\Psi}
\Pi_{\overline{\Psi}})
\label{cd9}
\end{equation}
and, on the constraint surface, commutes with all constraints. The
fermionic constraints are second class constraints and commute now with
all nonfermionic ones. Therefore we will omit them in future
consideration.

	The consistency condition of the constraint (\ref{cd9}) gives
neither  a new constraint nor determines any Lagrange multiplier.
The constraints $\phi^{(1,{\cal Q})}_{j}, \phi^{({\cal Q})}_{j+4},
\phi^{(1)}_{\Psi},\phi^{(1)}_{\overline{\Psi}}$ are second class, and
the $\phi^{(1,{\cal Q})}_{+}, \phi^{(2,{\cal Q})}_{+}$  are first class.
We need two gauge conditions. We choose them in the form
\begin{eqnarray}
& &\phi^{(1,{\cal Q})}_{G}={\cal Q}\ast A_{+},\nonumber \\
& &\phi^{(2,{\cal Q})}_{G}={\cal Q}\ast \partial_{-}A_{-}
\label{cd11}
\end{eqnarray}
So in the ${\cal Q}$-sector the constraints ${\bf \Phi}^{({\cal Q})} \approx
0$ are
\begin{equation}
{\bf \Phi}^{({\cal Q})}=\left\{ \begin{array}{ll}
\phi^{(1,{\cal Q})}_{+}\equiv \phi^{({\cal Q})}_{1}\\
\phi^{(1,{\cal Q})}_{G}\equiv \phi^{({\cal Q})}_{2}\\
-{\cal Q}\ast \Delta^{-1}_{-}(\partial_{-}\Pi_{-}+\partial_{k}\Pi_{k}+
ie(\Pi_{\Psi} \Psi+ \overline{\Psi}\Pi_{\overline{\Psi}}))\equiv
\phi^{({\cal Q})}_{3}\\
\phi^{(2,{\cal Q})}_{G}\equiv \phi^{({\cal Q})}_{4}\\
\phi^{(1,{\cal Q})}_{j}\equiv \phi^{({\cal Q})}_{j+4}\\
                             \end{array}
                             \right.
\label{cd10}
\end{equation}
where $\Delta_{-}=\partial^{2}_{-}$ and $\Delta^{-1}_{-}$ is a
nondegenerate operator
whose matrix elements  in the ${\cal Q}$-sector
are
\begin{eqnarray}
(\Delta^{-1}_{-})_{\underline{x},\underline{y}}&=&H^{{\cal Q}}(x^{-}-y^{-})
\delta(x_{\perp}-y_{\perp}), \nonumber \\
H^{{\cal Q}}(x-y)&=&\frac{|x-y|}{2}-\frac{(x-y)^{2}}{4L}
\label{cd10a}
\end{eqnarray}

	The constraints ${\bf \Phi}^{({\cal Q})}$ are second class.
The matrix of Poisson brackets between these constraints
\begin{equation}
A_{a,b}(\underline{x},\underline{y})=\{\phi^{({\cal Q})}_{a}(\underline{x}),
\phi^{({\cal Q})}_{b}(\underline{y})\},
\;a,b=1,\ldots ,6,
\label{cd12}
\end{equation}
is quasi-diagonal (block-diagonal) and its non-vanishing elements
are found to be
\begin{eqnarray}
A_{1,2}(\underline{x},\underline{y})=-A_{2,1}(\underline{x},\underline{y})
&=& -{\cal Q}(\underline{x},\underline{y}),\nonumber \\
A_{3,4}(\underline{x},\underline{y})=-A_{4,3}(\underline{x},\underline{y})
&=& -{\cal Q}(\underline{x},\underline{y}),\nonumber \\
A_{j+4,j^{\prime}+4}(\underline{x},\underline{y})
&=&-2\delta_{jj^{\prime}}\partial_{-}
{\cal Q}(\underline{x},\underline{y})
\end{eqnarray}
The inverse matrix $A^{-1}_{a,b}(\underline{y},\underline{z})$
\begin{eqnarray}
& &\int d\underline{y}A_{c,a}(\underline{x},\underline{y})A^{-1}_{a,b}
(\underline{y},\underline{z})=\delta_{cb}
\delta^{({\cal Q})}(\underline{x}-\underline{z}),\\
\label{cd13}
& &\delta^{({\cal Q})}(\underline{x}-\underline{z})
=D^{\cal Q}(x^{-}-z^{-})\delta(x_{\perp}-z_{\perp}),\nonumber \\
& &D^{\cal Q}(x-z)=\delta(x-z)-\frac{1}{2L},
\label{cd14}
\end{eqnarray}
has non-vanishing elements
\begin{eqnarray}
& &A^{-1}_{1,2}(\underline{x},\underline{y})=-A^{-1}_{2,1}(\underline{x},
\underline{y}) = -{\cal Q}(\underline{x},\underline{y}),\nonumber \\
& &A^{-1}_{3,4}(\underline{x},\underline{y})=-A^{-1}_{4,3}
(\underline{x},\underline{y})=-{\cal Q}(\underline{x},\underline{y}),
\nonumber \\
& &A^{-1}_{j+4,j^{\prime}+4}(\underline{x},\underline{y})
=-\frac{1}{2}\delta_{jj^{\prime}}
\delta(x_{\perp}-z_{\perp})G^{\cal Q}(x^{-}-z^{-})\\
\label{cd15}
& &G^{\cal Q}(x-z)=\frac{\epsilon(x-z)}{2}-\frac{x-z}{2L}
\label{cd16}
\end{eqnarray}
The functions $D^{\cal Q}(x-z)$ and $G^{\cal Q}(x-z)$
are the delta function and the matrix element of the operator
$\partial^{-1}_{-}$, respectively, in the $\cal Q$ sector.

	Consider two variables
\begin{equation}
\omega^{({\cal Q})}_{j}={\cal Q}\ast A_{j}
\label{cd17}
\end{equation}
It is easy to show that the Dirac brackets between these variables
are
\begin{equation}
\{\omega^{({\cal Q})}_{j}(\underline{x}),\omega^{({\cal Q})}_{j^{\prime}}
(\underline{y}\}_{D}=-\frac{1}{2}\delta_{j j^{\prime}}\delta(x^{\perp}-
y^{\perp})G^{{\cal Q}}(x^{-}-y^{-})
\label{cd18}
\end{equation}
Thus we can consider the variables $\omega^{({\cal Q})}_{j}(\underline{x})$
 as the physical variables
in the ${\cal Q}$-sector.

\setcounter{equation}{0}
\section{{\cal P}-sector}
\vskip2ex

\hspace {3em} In this Section we will consider the canonical
quantization procedure in the ${\cal P}$-sector. The
primary Hamiltonian density in this sector is
\begin{equation}
{\cal H}^{(1,{\cal P})}={\cal H}_{c}+
\lambda^{({\cal P})}_{+}\phi^{(1,{\cal P})}_{+}+\lambda^{({\cal P})}
_{j}\phi^{(1,{\cal P})}_{j}
\label{cd22}
\end{equation}

The ${\cal P}$-sector is a space $H^{({\cal P})}$ of functions which
depend on the variables $x^{\perp}$ only. On the other hand the ${\cal
P}$-sector is the sector of zero-modes in the $x^{-}$-direction. To take
into account the zero-mode contributions in the directions
$x^{1},x^{2}$  one can decompose the space $H^{({\cal P})}$
into a direct sum of two orthogonal subspaces
\begin{equation}
H^{({\cal P})}=H^{({\cal P}{\cal P}_{1}{\cal P}_{2})}\bigoplus
H^{({\cal P}({\cal Q}_{1}+{\cal Q}_{2}-{\cal Q}_{1}{\cal Q}_{2}))}
\label{1}
\end{equation}
where $H^{({\cal P}{\cal P}_{1}{\cal P}_{2})}$ is the space of
zero-modes in all directions and is defined by the projector
${\cal P}{\cal P}_{1}{\cal P}_{2}$ and the space
$H^{({\cal P}({\cal Q}_{1}+{\cal Q}_{2}-{\cal Q}_{1}{\cal Q}_{2}))}$
is defined by the projector
${\cal P}({\cal Q}_{1}+{\cal Q}_{2}-{\cal Q}_{1}{\cal Q}_{2})$. Such
a decomposition leads to the corresponding
decomposition of the primary constraints
$\phi^{(1,{\cal P})}_{+},\phi^{(1,{\cal P})}_{j}$ (\ref{cd6}),
and the primary
Hamiltonian  density ${\cal H}^{(1,{\cal P})}$ (\ref{cd22})
\begin{eqnarray}
& &\phi^{(1,{\cal P})}_{+}=\left (\varphi^{(1)}_{+},\;
\psi^{(1)}_{+} \right ),\nonumber \\
& &\phi^{(1,{\cal P})}_{j}=\left (\varphi^{(1)}_{j},\;
\psi^{(1)}_{j} \right ),\\
\label{2}
& &{\cal H}^{(1,{\cal P})}=
{\cal H}_{c}+ \lambda^{(1)}_{+}\varphi^{(1)}_{+}+
\Lambda^{(1)}_{+}\psi^{(1)}_{+}+
\lambda^{(1)}_{j}\varphi^{(1)}_{j}+
\Lambda^{(1)}_{j}\psi^{(1)}_{j}
\label{3}
\end{eqnarray}
Here $\lambda^{(1)}_{+}, \Lambda^{(1)}_{+}, \lambda^{(1)}_{j},
\Lambda^{(1)}_{j}$ are the Lagrange multipliers; the primary constraints
in the ${\cal P}{\cal P}_{1}{\cal P}_{2}$-subsector are
\begin{eqnarray}
& &\varphi^{(1)}_{+}={\cal P}{\cal P}_{1}{\cal P}_{2}\ast \Pi_{+},
\nonumber \\
& &\varphi^{(1)}_{j}={\cal P}{\cal P}_{1}{\cal P}_{2}\ast \Pi_{j}
\label{4}
\end{eqnarray}
and
\begin{eqnarray}
& &\psi^{(1)}_{+}={\cal P}({\cal Q}_{1}+{\cal Q}_{2}-{\cal Q}_{1}
{\cal Q}_{2})\ast \Pi_{+},\nonumber \\
& &\psi^{(1)}_{j}={\cal P}({\cal Q}_{1}+{\cal Q}_{2}-{\cal Q}_{1}
{\cal Q}_{2})\ast(\Pi_{j}+\partial_{j}A_{-}),
\label{5}
\end{eqnarray}
are the primary constraints
in the ${\cal P}({\cal Q}_{1}+{\cal Q}_{2}-
{\cal Q}_{1}{\cal Q}_{2})$-subsector.

The consistency condition for the constraints $\varphi^{(1)}_{+}
, \varphi^{(1)}_{j}$ and $\psi^{(1)}_{+}, \psi^{(1)}_{j}$
 do not determine their
Lagrange multipliers, and instead give new constraints in the form
(we denote them as $\chi^{(1)}_{+},\chi^{(1)}_{j},\chi^{(2)}_{+},
\chi^{(2)}_{j}$, correspondingly )
\begin{eqnarray}
& &\chi^{(1)}_{+}={\cal P}{\cal P}_{1}{\cal P}_{2}\ast(ie
\Pi_{\Psi}\Psi),
\label{cd23}\\
& &\chi^{(1)}_{j}={\cal P}{\cal P}_{1}{\cal P}_{2}\ast(
-ie\Pi_{\Psi} {U}^{-1}_{-}{U}_{j}\Psi)\\
\label{cd24}
& &\chi^{(2)}_{+}={\cal P}({\cal Q}_{1}+{\cal Q}_{2}-{\cal Q}_{1}
{\cal Q}_{2})\ast (\partial_{j}\Pi_{j}+
ie\Pi_{\Psi}\Psi), \\
\label{cd23.a}
& &\chi^{(2)}_{j}={\cal P}({\cal Q}_{1}+{\cal Q}_{2}-{\cal Q}_{1}
{\cal Q}_{2})\ast(\partial_{j}\Pi_{-}+
\partial_{k}F_{kj}-
ie\Pi_{\Psi}{U}^{-1}_{-}{U}_{j}\Psi)
\label{cd24a}
\end{eqnarray}

	As it happened in the ${\cal Q}$-sector the constraints
(\ref{cd23})-(\ref{cd24a}) do not commute with the fermionic constraints.
Thus let us consider new constraints which are equivalent to
(\ref{cd23})-(\ref{cd24a}) in the presence of (\ref{cd3a})
\begin{eqnarray}
& &\chi^{(1)}_{+}={\cal P}{\cal P}_{1}{\cal P}_{2}\ast \rho, \\
\label{cd25}
& &\chi^{(1)}_{j}={\cal P}{\cal P}_{1}{\cal P}_{2}\ast \rho_{j},\\
\label{cd26}
& &\chi^{(2)}_{+}={\cal P}({\cal Q}_{1}+{\cal Q}_{2}-{\cal Q}_{1}
{\cal Q}_{2})\ast (\partial_{j}\Pi_{j}+\rho),
\label{25a}\\
& &\chi^{(2)}_{j}={\cal P}({\cal Q}_{1}+{\cal Q}_{2}-{\cal Q}_{1}
{\cal Q}_{2})\ast(\partial_{j}\Pi_{-}+\partial_{k}F_{kj}-\rho_{j}),
\label{cd25b}
\end{eqnarray}
where
\begin{equation}
\rho=ie(\Pi_{\Psi}\Psi+\overline{\Psi}\Pi_{\overline{\Psi}}),
\label{26}
\end{equation}
and
\begin{equation}
\rho_{j}=\frac{U_{j}}{U_{-}}\rho,
\label{27}
\end{equation}
which is the consequence of the heavy mass limit.

Let us consider the quantization procedure where the strong constraint
\begin{equation}
{\cal P}{\cal P}_{1}{\cal P}_{2}\ast \rho =0
\label{28}
\end{equation}
holds. The quantity ${\cal P}{\cal P}_{1}{\cal P}_{2}\ast \rho $ is the
total light-cone electric charge of the fermions per unit volume. This
is proportional to the usual electric charge, which must vanish in a
compact system as a consequence of Gauss's law.

	The consistency condition for the constraints
(\ref{25a}),(\ref{cd25b}) does not give
any new constraints. So in the ${\cal P}$-sector we have fourteen
 constraints
: $\varphi^{(1)}_{+}$, $\varphi^{(1)}_{j}$,
 $\chi^{(1)}_{+}$, $\chi^{(1)}_{j}$, $\psi^{(1)}_{+}$, $\psi^{(1)}_{j}$,
  $\chi^{(2)}_{+}$, $\chi^{(2)}_{j}$, $\phi^{(1)}_{\Psi}$,
 $\phi^{(1)}_{\overline{\Psi}}$.
  The constraints $\varphi^{(1)}_{+}$, $\varphi^{(1)}_{j}$,
$\psi^{(1)}_{+}$ and $\chi^{(2)}_{+}$ are first class and the rest
are second class.  We need four gauge conditions.

	We consider the  gauge conditions for the first class constraints
$\varphi^{(1)}_{+}$ and $\psi^{(1)}_{+}$ in the form
\begin{eqnarray}
& &\varphi^{(1)}_{G}={\cal P}{\cal P}_{1}{\cal P}_{2}\ast A_{+},\nonumber
\\
& &\psi^{(1)}_{G}={\cal P}({\cal Q}_{1}+{\cal Q}_{2}-{\cal Q}_{1}
{\cal Q}_{2})\ast  A_{+}
\label{cd27}
\end{eqnarray}

Let us discuss the gauge conditions for the first class constraint
$\chi^{(2)}_{+}$ in the ${\cal P}({\cal Q}_{1}+{\cal
Q}_{2}-{\cal Q}_{1}{\cal Q}_{2})$-subsector. Note that in this subsector
the operator
\begin{equation}
\Delta=\partial_{k}\partial_{k}
\label{28a}
\end{equation}
is an invertible operator. Let us choose the gauge condition for the
constraints $\chi^{(2)}_{+}$ in the form
\begin{equation}
\chi^{(2)}_{G}={\cal P}({\cal Q}_{1}+{\cal Q}_{2}-{\cal Q}_{1}{\cal
Q}_{2})\ast \Delta ^{-1}\partial_{k}A_{k}
\label{28b}
\end{equation}

	The constraints $\chi^{(2)}_{j}$ do not commute with
the primary constraints $\psi^{(1)}_{j}$ but do commute with the
rest. On the other hand, the primary constraints $\psi^{(1)}_{j}$
do not commute with the gauge condition $\chi^{(2)}_{G}$.
Thus let us consider a linear combination of the constraints
$\psi^{(1)}_{j}$ and $\chi^{(2)}_{+}$
\begin{equation}
\psi^{\prime(1)}_{j}=\psi^{(1)}_{j}+ \alpha_{j}\chi^{(2)}_{+}
\label{28c}
\end{equation}
in such a way that $\psi^{\prime(1)}_{j}$ will commute with the
gauge condition $\chi^{(2)}_{G}$.  Here $\alpha_{j}$ are some
functions to be found. We obtain
\begin{eqnarray}
\psi^{\prime(1)}_{j}&=&{\cal P}({\cal Q}_{1}+{\cal Q}_{2}-{\cal
Q}_{1}{\cal Q}_{2})\ast(\Pi^{\perp}_{j} -
\frac{\partial_{j}}{\Delta}\rho +\partial_{j}A_{-}),\\
\label{cd29}
\Pi^{\perp}_{j}&=&(\delta_{jk}-\frac{\partial_{j}\partial_{k}}{\Delta})
\Pi_{k}
\label{cd30}
\end{eqnarray}
Here $\Pi^{\perp}_{j}$ are two-dimensional transverse momenta. Let us
introduce longitudinal momenta and an analogous decomposition for the
potentials $A_{k}$
\begin{eqnarray}
\Pi_{j}&=&\Pi^{\perp}_{j}+\Pi^{\parallel}_{j},\;\Pi^{\parallel}_{j}
=\frac{\partial_{j}\partial_{k}}{\Delta}\Pi_{k},\\
\label{cd31a}
A_{j}&=&A^{\perp}_{j}+A^{\parallel}_{j},\nonumber \\
A^{\perp}_{j}&=&(\delta_{jk}-\frac{\partial_{j}\partial_{k}}{\Delta})
A_{k},\;A^{\parallel}_{j}=\frac{\partial_{j}\partial_{k}}{\Delta}A_{k}
\label{cd32}
\end{eqnarray}

With this decomposition the constraints depend only on transverse
or longitudinal components, but not both
\begin{eqnarray}
& &\chi^{(2)}_{+}={\cal P}({\cal Q}_{1}+{\cal Q}_{2}-{\cal Q}_{1}
{\cal Q}_{2})\ast(\partial_{j}\Pi^{\parallel}_{j}+\rho),\\
\label{cd33}
& &\chi^{(2)}_{G}={\cal P}({\cal Q}_{1}+{\cal Q}_{2}-{\cal Q}_{1}
{\cal Q}_{2})\ast \Delta ^{-1}
\partial_{k}A^{\parallel}_{k},\\
\label{cd34aa}
& &\chi^{(2)}_{j}={\cal P}({\cal Q}_{1}+{\cal Q}_{2}-{\cal Q}_{1}
{\cal Q}_{2})\ast(\partial_{j}\Pi_{-}+
\Delta A^{\perp}_{j}-\rho_{j})
\label{cd35}
\end{eqnarray}

Consider the primary first class constraints $\varphi^{(1)}_{j}$. Naively
one might try to choose the gauge condition as
${\cal P}{\cal P}_{1}{\cal P}_{2} \ast A_{j} \approx 0$,
which means that the zero-modes of $A_{j}$ are
eliminated. This contradicts, however, the fact that the integral
over a closed loop
\begin{equation}
\oint A_{j}dx_{j}
\label{cd35a}
\end{equation}
is  gauge invariant \cite{Wilson}, and need not vanish if taken over a
non-contractible loop. Instead, we choose the gauge
condition $\varphi^{(1)}_{j G}\approx 0$ in the form
\begin{equation}
\varphi^{(1)}_{j G}={\cal P}{\cal P}_{1}{\cal P}_{2}\ast
A_{j}-f_{j}(x^{+}),
\label{cd35b}
\end{equation}
where $f_{j}(x^{+})$ is any function of the ``time'' $x^{+}$. The
constraints $\varphi^{(1)}_{j G}$ depend on ``time'' explicitly. It is
well known that the consistency condition for this type of constraint
can be written in the form
\begin{equation}
\partial_{+}\varphi^{(1)}_{j G}+\int\{ \varphi^{(1)}_{j G}, {\cal H}_
{c}(\underline{x}) \}d\underline{x} \approx 0
\label{cd35c}
\end{equation}
This relation determines the Lagrange multiplier $\lambda^{(1)}_{j}$ in
terms of the derivative $\partial_{+}f_{j}$. In the next Section we will
show that the physical Hamiltonian does not depend on the function
$f_{j}(x^{+})$.

So the constraints ${\bf \Phi}^{({\cal P})}\approx 0$ in the
 ${\cal P}$ sector are (we omit here the fermionic constraints)
\begin{equation}
{\bf \Phi}^{({\cal P})}= \left\{ \begin{array}{ll}
            \left.\begin{array}{ll}
\varphi^{(1)}_{+}\equiv \phi^{({\cal P})}_{1} \\
\varphi^{(1)}_{G}\equiv \phi^{({\cal P})}_{2} \\
\psi^{(1)}_{+}\equiv \phi^{({\cal P})}_{3} \\
\psi^{(1)}_{G}\equiv \phi^{({\cal P})}_{4} \\
\varphi^{(1)}_{j}\equiv \phi^{({\cal P})}_{j+4} \\
\varphi^{(1)}_{j G}\equiv \phi^{({\cal P})}_{j+6}\end{array}\right\}
\mbox{${\cal P}{\cal P}_1{\cal P}_2$-subsector}\\
\left.\begin{array}{ll}
\chi^{(2)}_{+}\equiv \phi^{({\cal P})}_{9}\\
\chi^{(2)}_{G}\equiv \phi^{({\cal P})}_{10}\\
\psi^{\prime(1)}_{j}\equiv \phi^{({\cal P})}_{j+10} \\
\chi^{(2)}_{j}\equiv \phi^{({\cal P})}_{j+12}
                       \end{array}\right\}\mbox{${\cal P}({\cal Q}_1+
{\cal Q}_2-{\cal Q}_1{\cal Q}_2)$-subsector}
                       \end{array}\right.
\label{cd36}
\end{equation}
The constraints (\ref{cd36}) are second-class constraints and the
matrix of their Poisson brackets is block-diagonal.

	The constraints $\phi^{({\cal P})}_{a},\;(a=1,\cdots ,10)$,
determine the nonphysical variables. The physical variables are among
the constraints $\phi^{({\cal P})}_{j+10},\phi^{({\cal P})}_{j+12}$. The
matrix of Poisson brackets between these constraints is antidiagonal and
its nonzero elements are
\begin{eqnarray}
\{ \phi^{({\cal P})}_{j+10}(\underline{x}),\phi^{({\cal
P})}_{j^{\prime}+12}(\underline{y}) \}&=&A_{j+10,j^{\prime}+12}(\underline{x},
\underline{y})\nonumber \\
& &=-\delta_{jj^{\prime}}\Delta ({\cal P}
({\cal Q}_{1}+{\cal Q}_{2}-{\cal Q}_{1}{\cal Q}_{2}))
(\underline{x},\underline{y}),\\
\label{cd37}
A_{j+12,j^{\prime}+10}(\underline{x},\underline{y})
&=&\delta_{jj^{\prime}}\Delta ({\cal P}
({\cal Q}_{1}+{\cal Q}_{2}-{\cal Q}_{1}{\cal Q}_{2}))
(\underline{x},\underline{y})
\label{cd38}
\end{eqnarray}
The inverse matrix is antidiagonal too, and it easily found to be
\begin{eqnarray}
& &A^{-1}_{j+12,j^{\prime}+10}(\underline{x},\underline{y})
=-\delta_{jj^{\prime}}
\frac{1}{\Delta}({\cal P}({\cal Q}_{1}+{\cal Q}_{2}-{\cal Q}_{1}
{\cal Q}_{2}))(\underline{x},\underline{y}),\nonumber \\
& &A^{-1}_{j+10,j^{\prime}+12}(\underline{x},\underline{y})
=\delta_{jj^{\prime}}\frac{1}{\Delta}({\cal P}
({\cal Q}_{1}+{\cal Q}_{2}-{\cal Q}_{1}{\cal Q}_{2}))
(\underline{x},\underline{y})
\label{cd39}
\end{eqnarray}

	Consider two pairs of conjugated variables
\begin{eqnarray}
& &\omega^{({\cal P})}_{1}={\cal P}({\cal Q}_{1}+{\cal Q}_{2}-
{\cal Q}_{1}{\cal Q}_{2})\ast A_{-},\nonumber \\
& &\omega^{({\cal P})}_{2}={\cal P}({\cal Q}_{1}+{\cal Q}_{2}-
{\cal Q}_{1}{\cal Q}_{2})\ast \Pi_{-}\\
\label{cd40}
& & \omega^{({\cal P})}_{3}={\cal P}{\cal P}_{1}{\cal P}_{2}\ast A_{-},
 \nonumber \\
& & \omega^{({\cal P})}_{4}={\cal P}{\cal P}_{1}{\cal P}_{2}\ast \Pi_{-}
\label{cd40a}
\end{eqnarray}

The non-zero Dirac brackets between these variables are
\begin{eqnarray}
\{ \omega^{({\cal P})}_{1}(\underline{x}),\omega^{({\cal P})}
_{2}(\underline{y}) \}_{D}&=&2({\cal P}({\cal Q}_{1}+{\cal Q}_{2}-
{\cal Q}_{1}{\cal Q}_{2}))(\underline{x},\underline{y}) \nonumber \\
& &=\frac{1}{2L}(\delta(x_{\perp}-y_{\perp})-\frac{1}{4L^2}),\\
\label{cd41}
\{ \omega^{({\cal P})}_{3}(\underline{x}),\omega^{({\cal P})}
_{4}(\underline{y}) \}_{D}&=&({\cal P}{\cal P}_{1}{\cal P}_{2})
(\underline{x},\underline{y})=\frac{1}{8L^{3}}
\label{cd41a}
\end{eqnarray}
Thus one can consider these variables as physical variables in the
${\cal P}$-sector.
\setcounter{equation}{0}
\section{Physical Hamiltonian}
\vskip2ex

\hspace {3em}We have come to the step where we can find the``physical''
 Hamiltonian. Such a Hamiltonian is defined as
\begin{eqnarray}
& &H^{phy}=\int {\cal H}^{phy}_{c}d \underline{x},
\nonumber \\
& &{\cal H}^{phy}_{c}={\cal H}_{c}|_{{\bf \Phi}\approx 0}
\label{cd19}
\end{eqnarray}
where
\begin{equation}
{\bf \Phi}=({\bf \Phi}^{({\cal Q})},\;{\bf \Phi}^{({\cal P})},\;
\phi^{(1)}_{\Psi},\;\phi^{(1)}_{\overline{\Psi}})
\label{cd19a}
\end{equation}
are all constraints in the problem under consideration.

	From  eq.(\ref{cd19}) it follows that
to construct the Hamiltonian density ${\cal H}^{phy}_{c}$ we have
to express all variables through the physical variables using
the constraints ${\bf \Phi}\approx 0$.
 For this purpose let us return to the
expression (\ref{cd2p}) for the canonical Hamiltonian
 and rewrite it in the following form
\begin{equation}
H_{c}=\int {\cal H}^{(F)}_{c}d\underline{x}+\int {\cal H}^{\prime}_{c}
d\underline{x}
\label{cd48}
\end{equation}
Here
\begin{equation}
{\cal H}^{(F)}_{c}=\overline{\Psi}\uslash (-i{U}_{+}\partial_{-}+
i{U}_{k}\partial_{k})\Psi+M\overline{\Psi}\Psi
\label{cd49}
\end{equation}
is the Hamiltonian density of fermions. The expression ${\cal H}
^{\prime}_{c}$
\begin{eqnarray}
& &{\cal H}^{\prime}_{c}= \frac{1}{2}({\cal Q}\ast\Pi_{-})^{2}+
\frac{1}{2}({\cal P}{\cal P}_{1}{\cal P}_{2}\ast\Pi_{-}+{\cal P}(
{\cal Q}_{1}+{\cal Q}_{2}-{\cal Q}_{1}{\cal Q}_{2})\ast \Pi_{-})^{2}
\nonumber \\
& &+({\cal Q}\ast\Pi_{-})({\cal P}{\cal P}_{1}{\cal P}_{2}\ast\Pi_{-}+
{\cal P}({\cal Q}_{1}+{\cal Q}_{2}-{\cal Q}_{1}{\cal Q}_{2})\ast
\Pi_{-})- \nonumber \\
& &(\partial_{-}\Pi_{-}+\partial_{j}\Pi_{j})({\cal Q}\ast A_{+}+
{\cal P}{\cal P}_{1}{\cal P}_{2}\ast A_{+}+
{\cal P}({\cal Q}_{1}+{\cal Q}_{2}-{\cal Q}_{1}{\cal Q}_{2})\ast A_{+})
\nonumber \\
& &+\frac{1}{4}({\cal Q}\ast F_{jk})^{2}
+\frac{1}{4}({\cal P}({\cal Q}_{1}+{\cal Q}_{2}-{\cal Q}_{1}{\cal
Q}_{2})\ast F_{jk})^{2}+\nonumber \\
& &\frac{1}{2}({\cal Q}\ast F_{jk})
({\cal P}({\cal Q}_{1}+{\cal Q}_{2}-{\cal Q}_{1}{\cal Q}_{2})\ast F_{jk})
\nonumber \\
& &+e\overline{\Psi}\uslash {U}^{\mu}\Psi
({\cal Q}\ast A_{\mu}+{\cal P}{\cal P}_{1}{\cal P}_{2}\ast A_{\mu}+
{\cal P}({\cal Q}_{1}+{\cal Q}_{2}-{\cal
Q}_{1}{\cal Q}_{2})\ast A_{\mu})
\label{cd50}
\end{eqnarray}
is the Hamiltonian density of the electromagnetic field and its
interaction with the fermions.
Then the physical Hamiltonian becomes
\begin{eqnarray}
& &H^{phy}=\int {\cal H}^{F}_{c}d \underline{x}+\int {\cal H}^{\prime,
phys}_{c}d\underline{x},
\nonumber \\
& &{\cal H}^{\prime,phy}_{c}={\cal H}^{\prime}_{c}|_{{\bf \Phi}\approx 0}
\label{cd52}
\end{eqnarray}

Due to the property
\begin{equation}
e\overline \Psi \uslash {U}^{\mu}\Psi=U^{\mu}\rho
\label{cd52a}
\end{equation}
and the strong condition (\ref{28}) one can neglect the term
$e\overline \Psi \uslash {U}^{\mu}\Psi{\cal P}{\cal P}_{1}{\cal P}_{2}\ast
A_{\mu}$ in the Hamiltonian density ${\cal H}^{\prime,phys}_{c}$
(\ref{cd52}). This is because the physical Hamiltonian $H^{\prime,phys}$
, which corresponds to the density ${\cal H}^{\prime,phys}_{c}$, by
itself is proportional to the zero-mode of
${\cal H}^{\prime,phys}_{c}$
\begin{equation}
H^{\prime,phys}=8L^{3}{\cal P}{\cal P}_{1}{\cal P}_{2}\ast
{\cal H}^{\prime,phys}_{c}
\label{cd52b}
\end{equation}
Thus the physical Hamiltonian does not depend on the functions
$f_{j}(x^{+})$ involved in the gauge condition (\ref{cd35b}).

	Using the constraints ${\bf \Phi}\approx 0$ we are able to write
\begin{eqnarray}
& &{\cal Q}\ast \Pi_{-}|_{{\bf \Phi}\approx 0}=-(\partial_{j}\omega
^{({\cal Q})}_{j}+e{\cal Q}\ast
\partial^{-1}_{-}\overline{\Psi}\uslash {U}_{-}\Psi ), \nonumber \\
& &{\cal Q}\ast A_{\pm}|_{{\bf \Phi}\approx 0}=0, \nonumber \\
& &{\cal Q}\ast F_{jk}|_{{\bf \Phi}\approx 0}=\partial_{j}\omega
^{({\cal Q})}_{k}-\partial_{k}\omega^{({\cal Q})}_{j}, \nonumber \\
& &{\cal P}{\cal P}_{1}{\cal P}_{2}\ast A_{+}|_{{\bf \Phi}\approx 0}=0,
\nonumber \\
& &{\cal P}({\cal Q}_{1}+{\cal Q}_{2}-{\cal Q}_{1}{\cal Q}_{2})\ast
A_{+}|_{{\bf \Phi}\approx 0}=0,\nonumber \\
& &{\cal P}{\cal P}_{1}{\cal P}_{2}\ast A_{j}|_{{\bf \Phi}\approx 0}=
f_{j}(x^{+}),\nonumber \\
& &{\cal P}({\cal Q}_{1}+{\cal Q}_{2}-{\cal Q}_{1}{\cal Q}_{2})
\ast A^{\parallel}_{k}|_{{\bf \Phi}\approx 0}=0, \nonumber \\
& &{\cal P}({\cal Q}_{1}+{\cal Q}_{2}-{\cal Q}_{1}{\cal Q}_{2})
\ast A^{\perp}_{k}|_{{\bf \Phi}\approx 0}=
-\frac{\partial_{k}}{\Delta}\omega^{({\cal P})}_{2}+\frac{U_{k}}{U_{-}}
\omega^{({\cal P})}_{1},\nonumber \\
& &{\cal P}({\cal Q}_{1}+{\cal Q}_{2}-{\cal Q}_{1}{\cal Q}_{2})
\ast F_{jk}|_{{\bf \Phi}\approx 0}=
{\cal P}({\cal Q}_{1}+{\cal Q}_{2}-{\cal Q}_{1}{\cal Q}_{2})
\ast F^{\perp}_{jk}|_{{\bf \Phi}\approx 0}\nonumber \\
& &=U^{-1}_{-}(U_{k}\partial_{j}-U_{j}\partial_{k})
\omega^{({\cal P})}_{1},
\label{cd53}
\end{eqnarray}

	After substitution of eqs.(\ref{cd53}) into eq.(\ref{cd52})
 we get
\begin{eqnarray}
{\cal H}^{\prime, phy}_{c}&=&\frac{1}{2}(\partial_{j}\omega ^{({\cal
Q})}_{j}+e{\cal Q}\ast\partial^{-1}_{-}\overline{\Psi}\uslash {U}_{-}\Psi-
\omega^{({\cal P})}_{2}-\omega^{({\cal P})}_{4})^{2}\nonumber \\
& &+\frac{1}{4}({\cal F}^{({\cal Q})}_{jk}+{\cal F}^{({\cal P})}_{jk})
({\cal F}^{({\cal Q})}_{jk}+{\cal F}^{({\cal P})}_{jk})\nonumber \\
& &-U^{-1}_{-}e\overline{\Psi}\gamma \cdot U\Psi U_{j}
\omega ^{({\cal Q})}_{j}+e\overline{\Psi}\uslash\Psi \frac{U_{j}
\partial_{j}}{U_{-}\Delta}\omega^{({\cal P})}_{2}\nonumber \\
& &+\frac{1-U_{-}U_{+}}{U^{2}_{-}}e\overline{\Psi}\uslash\Psi
\omega^{({\cal P})}_{1}
\label{cd54a}
\end{eqnarray}
where
\begin{eqnarray}
& &{\cal F}^{({\cal Q})}_{jk}=\partial_{j}\omega ^{({\cal Q})}_{k}-
\partial_{k}\omega ^{({\cal Q})}_{j},\nonumber \\
& &{\cal F}^{({\cal P})}_{jk}=
U^{-1}_{-}(U_{k}\partial_{j}-U_{j}\partial_{k})
\omega^{({\cal P})}_{1}
\label{cd54b}
\end{eqnarray}

Having the physical Hamiltonian and the Poisson brackets between
all physical variables
one can now  quantize  heavy fermion QED.
According to the Dirac prescription the quantization procedure consists of
replacing  the Dirac brackets of bosons with bosons (or bosons
with fermions) by a commutator
and that of fermions with fermions by an
anticommutator
\begin{eqnarray}
& &\{A,B\}_{D}\rightarrow -i[\hat{A},\hat{B}]
\mbox{-for  bosons},\nonumber \\
& &\{A,B \}_{D}\rightarrow -i[ \hat{A},\hat{B} ]_+ \mbox{-for fermions}
\label{55}
\end{eqnarray}
where the symbol $\;\hat{}\;$  means that the classical object $A$ becomes an
operator $\hat{A}$.
Then we come to the following equal ``time'' commutation
relations
\begin{eqnarray}
& &[\hat{\omega}^{({\cal Q})}_{j}(\underline{x}),\hat{\omega}^{({\cal Q})}
_{k}(\underline{y})]=-\frac{i}{2}\delta_{jk}\delta(x_{\perp}-y_{\perp})
G(x^{-}-y^{-}),
\label{56a}\\
& &\left [\hat{\omega}^{({\cal P})}_{1}(x_{\perp}),\hat{\omega}^{({\cal
P})}_{2}
(y_{\perp})\right
]=\frac{i}{L}(\delta(x_{\perp}-y_{\perp})-\frac{1}{4L^{2}}),\\
\label{56b}
& &[\hat{\omega}^{({\cal P})}_{3},\hat{\omega}^{({\cal P})}_{4}]=
\frac{i}{8L^{3}},\\
\label{56c}
& & [ \hat{\Psi}(\underline{x}), \hat{\overline{\Psi}}(\underline{y})
]_+=\frac{\uslash}{U_{-}}\delta(\underline{x}-\underline{y})
\label{56d}
\end{eqnarray}
The physical Hamiltonian $\hat{H}^{phys}$ can be obtained from (\ref{cd54a})
by the change $\omega_{a} \rightarrow \hat{\omega}_{a}$, where
$\hat{\omega}_a$
are all operators satisfying the relations (\ref{56a})-(\ref{56d}).

We wish to find a realization of the commutation relations
(\ref{56a})-(\ref{56d}). Using the periodic boundary conditions for
$\omega^{({\cal Q})}_{j} (\underline{x})$ and eq.(\ref{56a}) one may
write (from now on we will omit the operator symbol $\;\hat{}\;$ )
\begin{equation}
\omega^{({\cal Q})}_{j}(\underline{x})=\frac{1}{\sqrt{\Omega}}\sum
_{\underline {P}_{n}}\frac{1}{\sqrt{2P^{+}_{n}}}\left (
a_{j}(\underline{P}_{n})\exp (-i\underline{P}_{n}\cdot \underline{x})
+a^{+}_{j}(\underline{P}_{n})\exp(i\underline{P}_{n}\cdot \underline{x})
\right ),
\label{57}
\end{equation}
where
\begin{equation}
\underline{P}_{n}=\frac{\pi}{L}(n^{+},n^{j}),\; n^{+}=1,2,\ldots ,\;
n^{j}=0,\pm1,\pm2,\ldots,\;\Omega=8L^{3}
\label{58}
\end{equation}
and for any vectors $\underline{a}\cdot
\underline{b}=a^{+}b^{-}-a_{j}b_{j}$.
The operators $a_{j}(\underline{P}_{n})$ and
$a^{+}_{j}(\underline{P}_{n})$ satisfy the following nonzero commutation
relations
\begin{equation}
\left [a_{j}(\underline{P}_{n}),a^{+}_{j^{\prime}}(\underline{P}_{m})
\right ]=i\delta_{jj^{\prime}}\delta_{\underline{P}_{n}\underline{P}_{m}}
\label{59}
\end{equation}

Consider the operators $\omega^{({\cal P})}_{1,2}(x_{\perp})$. Their
decomposition into a Fourier series can be written in the following form
\begin{eqnarray}
\omega^{({\cal P})}_{1}(x_{\perp})&=&\frac{1}{\sqrt{\Omega}}
\sum_{P_{\perp n}}(a(P_{\perp n})\exp(-iP_{\perp n}\cdot x_{\perp})+
a^{+}(P_{\perp n})\exp(iP_{\perp n}\cdot x_{\perp}))\nonumber \\
& &+\frac{i}{\sqrt{\Omega}}(a-a^{+}),\nonumber \\
\omega^{({\cal P})}_{2}(x_{\perp})&=&\frac{i}{\sqrt{\Omega}}
\sum_{P_{\perp n}}(a(P_{\perp n})\exp(-iP_{\perp n}\cdot x_{\perp})-
a^{+}(P_{\perp n})\exp(iP_{\perp n}\cdot x_{\perp}))\nonumber \\
& &-\frac{1}{\sqrt{\Omega}}(a+a^{+}),
\label{60}
\end{eqnarray}
where
\begin{equation}
P_{\perp n}=\frac{\pi}{L}n^{j},\;n^{j}=0,\pm1,\pm2,\ldots,\;
a_{\perp}\cdot b_{\perp}=a_{j}b_{j},
\label{60a}
\end{equation}
and the operators $a(P_{\perp n}),a^{+}(P_{\perp n}),a,a^{+}$
satisfy the following nonzero commutation relations
\begin{eqnarray}
& &\left [a(P_{\perp n}),a^{+}(P_{\perp m})\right ]
=\delta_{P_{\perp n}P_{\perp m}},\nonumber \\
& &[a,a^{+}]=1
\label{61}
\end{eqnarray}
The realization for the zero-mode operators $\omega^{({\cal P})}_{3,4}$
is
\begin{eqnarray}
& &\omega^{({\cal P})}_{3}=\frac{1}{\sqrt{\Omega}}(c+c^{+});\;
\omega^{({\cal P})}_{4}=-\frac{i}{\sqrt{\Omega}}(c-c^{+}),\nonumber \\
& &[c,c^{+}]=1
\label{62}
\end{eqnarray}
The operators $a_{j}(\underline{P}_{n}),a(P_{\perp n}),a,c$ and
$a^{+}_{j}(\underline{P}_{n}),a^{+}(P_{\perp n}),a^{+},c^{+}$ can be
interpreted as annihilation and creation operators.

The same decomposition in terms of creation and annihilation operators
can be made for the fermion operators $\Psi(\underline{x}),\overline{\Psi}
(\underline{x})$ satisfying the antiperiodic boundary conditions
\begin{equation}
\Psi(\underline{x})=\frac{1}{\sqrt{\Omega U_{-}}}
\sum_{\beta,\underline{Q_{n}}}(b_{\beta}(\underline{Q_{n}})
u^{\beta}_{+}\exp(-i\underline{Q_{n}}\cdot\underline{x})
+d^{+}_{\beta}(\underline{Q_{n}})u^{\beta}_{-}
\exp(i\underline{Q_{n}}\cdot\underline{x}))
\label{63}
\end{equation}
where
\begin{eqnarray}
& & \underline{Q_{n}}=(Q^{+}_{n},Q^{j}_{n})=
\frac{\pi}{L}(n^{+}+1/2,n^{j}+1/2),\nonumber \\
& &n^{+}=0,1,2,\cdots ,\;n^{j}=0,\pm1,\pm2,\cdots \nonumber \\
& &\Lambda_{\pm}u^{\beta}_{\pm}=u^{\beta}_{\pm},\;
\Lambda_{\pm}=\frac{1}{2}(1\pm \uslash),\nonumber \\
& &\bar{u}^{\beta}_{\pm}u^{\alpha}_{\pm}=\pm \delta_{\alpha \beta},\;
\bar{u}^{\beta}_{\pm}u^{\alpha}_{\mp}=0,
\label{64}
\end{eqnarray}
and the operators $b_{\beta}(\underline{Q}_{n}),
d^{+}_{\beta}(\underline{Q}_{n})$ satisfy the following nonzero
commutation relations
\begin{equation}
[ b_{\beta}(\underline{Q}_{n}), b^{+}_{\alpha}
(\underline{Q}_{m}) ]_{+}=
[ d_{\beta}(\underline{Q}_{n}), d^{+}_{\alpha}(\underline{Q}_{m})]_{+}
=\delta_{\alpha \beta}\delta_{\underline{Q}_{n} \underline{Q}_{m}}
\label{65}
\end{equation}

\setcounter{equation}{0}
\section{Calculation of Heavy Fermion Potential}
\vskip2ex

\hspace{3em}We wish to see the physical Hamiltonian $H^{phys}$
found in the previous Section in action.  The example we
choose is the static heavy fermion potential. (This calculation is the
analog in QED of the calculation of the quark-antiquark potential
in QCD.)
For this purpose we consider old-fashioned perturbation theory
up to second order
\begin{equation}
E\approx E_{(0)}+eE_{(1)}+e^{2}E_{(2)}
\label{cd26aa}
\end{equation}

The physical Hamiltonian $H^{phys}$ can be expressed in terms of the
normal product of the creation and annihilation operators introduced in
Section 5
\begin{equation}
H^{phys}(operator)\equiv H_{(0)}+eH_{(1)}+e^{2}H_{(2)}
\label{cd30a}
\end{equation}

The construction of the normalized eigenstate which describes the two static
heavy fermions $\underline{r}$ apart yields
\begin{equation}
|f\bar{f}\rangle=(\sum_{\underline{Q}_{m}}^{M}1)^{-1/2}\sum_{\underline{Q}
_{n}}
^{M}\exp(-i(\underline{Q}_{n}-M\underline{U})\cdot\underline{r})
b^{+}_{\alpha}(\sqrt{2}M-Q^{+}_{n},-Q_{\perp n})
d^{+}_{\beta}(\underline{Q}_{n})|0\rangle
\label{cd31}
\end{equation}
where $\alpha\neq\beta$ and we choose
\begin{eqnarray*}
U^{+}=U^{-}=\frac{1}{\sqrt{2}},\;U^{1}=U^{2}=0
\end{eqnarray*}
In eq.(\ref{cd31}) we have the effective summation over
$\underline{Q}_{n}$ which has modulus less than $M$. It can
be shown that the corresponding eigenvalue
$(E_{(0)})$ of $H_{(0)}$ turns out to be $\sqrt{2}M$ and the mean value
of the operator $H_{(1)}$ in the state (\ref{cd31}) is zero
\begin{equation}
E_{(1)}=
\langle \bar{f}f|H_{(1)}|f\bar{f}\rangle=0
\label{cd311}
\end{equation}

The last term in eq.(\ref{cd26aa}) is
\begin{equation}
E_{(2)}=
\langle \bar{f}f|H_{(2)}|f\bar{f}\rangle+
\sum_{m}'\frac{|\langle \bar{f}f|H_{(1)}|m\rangle|^{2}}
{E_{(0)}-E^{m}_{(0)}}
\label{cd312}
\end{equation}
where $|m\rangle$ is an eigenstate of $H_{(0)}$ and $E^{m}_{(0)}$
is the corresponding eigenvalue and the summation does not cover the
state $|f\bar{f}\rangle$. After some straightforward calculation we get
\begin{equation}
\langle \bar{f}f|H_{(2)}|f\bar{f}\rangle=-\frac{1}{2\Omega}\sum_{\underline{
P}_{n}}\frac{1}{(P^{+}_{n})^{2}}[1+\cos(\underline{P}_{n}\cdot\underline{r})]
\label{cd32a}
\end{equation}
and
\begin{equation}
\sum_{m}'\frac{|\langle \bar{f}f|H_{(1)}|m\rangle|^{2}}
{E_{(0)}-E^{m}_{(0)}}=\frac{1}{2\Omega}\sum_{\underline{P_{n}}}
\frac{1}{2P^{+\;2}_{n}+P^{2}_{\perp n}}\frac{P^{2}_{\perp n}}{P^{+\;2}_{n}}
[1+\cos(\underline{P}_{n}\cdot\underline{r})]             \label{cd33aa}
\end{equation}

If we remember that $ \underline{r}=(-\frac{1}{\sqrt{2}}r^{3},r^{1},r^{2})$ and
make use of the new vectors
\begin{equation}
{\bf P}_{n} \equiv (P^{1}_{n},P^{2}_{n},\sqrt{2}P^{+}_{n}),{\bf r} \equiv
(r^{1},r^{2},r^{3}_{c})
\label{cd33a}
\end{equation}
then (in the limit $L \to \infty$),
\begin{equation}
e^{2}E_{(2)}=-\frac{e^{2}}{2\Omega}\sum_{{\bf P}_{n}}\frac{1}{{\bf P}_{n}^{2}}
[1+\cos({\bf P}_{n}\cdot{\bf r})]=-\frac{e^{2}}{4\sqrt{2}\pi r}+\mbox{Const},
\label{cd34}
\end{equation}
where $r^{3}_{c}$ is the Cartesian $z$-component of the vector ${\bf r}$.
The additional factor $1/\sqrt{2}$ in the expression (\ref{cd34}) comes
from the fact that the energy $e^{2}E_{(2)}$ corresponds to the
Hamiltonian $P^{-}$ in the light-cone coordinates. In the reference
frame where ${\bf P}=0$ we have
\begin{eqnarray*}
P^{-}=\frac{1}{\sqrt{2}}P^{0}
\end{eqnarray*}
This gives the potential in the Cartesian coordinates
\begin{equation}
V(r)=-\frac{e^{2}}{4\pi r}
\end{equation}
and clearly shows the Coulomb potential together with
an irrelevant infinite
constant in eq.(\ref{cd34}) (which is proportional to M) arising from
the fermion self energy.

\section{Conclusions}
\setcounter{equation}{0}

\hspace {3em} We have considered the light-cone canonical quantization
of the heavy fermion QED  taking into account the zero-mode
contributions explicitly. We have imposed periodic boundary
conditions for bosonic fields and antiperiodic ones for fermionic
fields. As in ordinary QED, this model is gauge invariant, which
means that there are unphysical degrees of freedom.  To quantize the
theory,  we have used the Dirac algorithm for a system with first-
and second-class constraints and the corresponding gauge conditions.
In order to make explicit the role of the zero-modes, we have
considered gauge fixing and quantization procedures in the zero-mode
and non-zero-mode sectors, separately. In all sectors we obtained the
physical variables and their canonical (anti-)commutation relations.
The physical Hamiltonian was constructed by excluding  unphysical
degrees of freedom.  We have considered the role of all physical fields
in the calculation of the potential between static heavy fermions.

We suggest that this approach can be used for the case of finite-mass
QED or, perhaps, one might first address heavy quark QCD.
An important goal is the calculation of the  QCD quark-antiquark potential.

\vskip3ex
\newpage
{\bf ACKNOWLEDGMENT}

\vskip3ex
	One of us (J.W.Jun) would like to thank the hospitality of
the members of the CWRU physics department during his visit.
We also thank Labros Petropoulos and Michael Thompson for assistance
in preparation of the manuscript. We are grateful to the National
Science Foundation for support of this work.


\vfill
\newpage
\begin{thebibliography}{99}

\bibitem{Isgur}
 N. Isgur and M. B. Wise, Phys.Lett.{\bf B232}, 113 (1989)

\bibitem{Eichten}
 E. Eichten and B. Hill, Phys.Lett.{\bf B234}, 511 (1990)

\bibitem{Georgi}
 H. Georgi, Phys.Lett.{\bf B240}, 447 (1990)

\bibitem{grev}
H. Georgi, ``Heavy Quark Effective Theory", Boulder TASI 91:589-630.

\bibitem{brev1}
S. J. Brodsky, et. al., ``The Challenge of light cone quantization of
gauge field theory", SLAC-PUB-5811 (June 1992).

\bibitem{brev2}
S. J. Brodsky and H. C. Pauli, ``Light cone quantization of quantum
chromodynamics", SLAC-PUB-5558 (June 1991).

\bibitem{Brodsky}
 S. J. Brodsky, T. Huang, G. Lepage and P. Mackenzie, Banff Summer Institute
on Particle Physics, Banff, Alberta, Canada, report CLNS-82/522 (1981)

\bibitem{Brodsky1}
 S. J. Brodsky and C. R. Ji, Application of QCD to hadrons and nuclear
interactions, in: Lecture Notes in Physics, {\bf 248}, ed.
C.A.Engelbrecht (Springer, Berlin, 1986)

\bibitem{Pauli}
 H. C. Pauli and S. J. Brodsky, Phys.Rev.{\bf D32}, 1993, 2001 (1985)

\bibitem{Pauli1}
 T. Eller, H. C. Pauli and S. J. Brodsky, Phys.Rev.{\bf D35}, 1493 (1987)

\bibitem{Pauli2}
 K. Hornbostel, H. C. Pauli and S. J. Brodsky, Phys.Rev.{\bf D41}, 3814
(1990)

\bibitem{Pauli3}
 R. J. Perry, A. Harindranath and K. G. Wilson,Phys.Rev.Lett.{\bf 65},
 2959 (1990)

\bibitem{Pauli4}
 R. J. Perry and A. Harindranath, Phys.Rev.{\bf D43}, 492, 4051 (1991)

\bibitem{mccartor}G. McCartor, Z. Phys. {\bf C 52}, 611 (1991).

\bibitem{Heinzl}
 Th. Heinzl, St. Krusche and E. Werner, Phys.Lett.{\bf B256}, 55 (1991);

\bibitem{Heinzl2}
 Th. Heinzl, St. Krusche and E. Werner, Phys. Lett. {\bf B275}, 410 (1992).

\bibitem{Maskawa}
 T. Maskawa and K. Yamawaki, Progr.Theor.Phys.{\bf 56}, 270 (1976)

\bibitem{Tang}
 A. C. Tang, S. J. Brodsky and H. C. Pauli, Phys.Rev.{\bf D44}, 1842 (1991)

\bibitem{Mustaki}
 D. Mustaki, Phys.Rev.{\bf D42}, 1184 (1990)

\bibitem{burkardt} M. Burkardt and E. Swanson, Phys. Rev. {\bf D46},
5083 (1992).

\bibitem{Dirac}
P. A. M. Dirac, Canad.J.Math.{\bf 2}, 129 (1950); Lectures on Quantum
 Mechanics (Benjamin, New York, 1964);

\bibitem{Bergmann}
 P. G. Bergmann, Helv.Phys.Acta, Suppl.{\bf 4}, 79 (1956)

\bibitem{Casalbuoni}
 R. Casalbuoni, Nuovo Cim. {\bf 33A}, 115, (1976)

\bibitem{Sundermayer}
 K. Sundermayer, Constrained Dynamics, Lecture Notes in Phys, {\bf 168},
(Springer, Berlin, 1982)

\bibitem{Gitman}
 D. M. Gitman and I. V. Tyutin, Quantization of Fields with Constraints
(Springer, Berlin, 1991)

\bibitem{Bardakci}
 K. Bardakci and M. B. Halpern, Phys.Rev.{\bf 175}, 1686, (1968)

\bibitem{Bardakci1}
 J. B. Kogut and D. E. Soper, Phys.Rev.{\bf D1}, 2901, (1970)

\bibitem{Bardakci2}
 F. Rohrlich, Acta Phys. Ausrt. {\bf 32}, 87, (1970)

\bibitem{Bardakci3}
 J. D. Bjorken, J. B. Kogut and D. E. Soper, Phys.Rev.{\bf D3}, 1382,(1971)

\bibitem{mccartor2}
G. McCartor, Z. Phys. {\bf C41}, 271 (1988).

\bibitem{Wilson}
K. G.  Wilson, Phys. Rev. {\bf D10}, 2445 (1976).

\end{thebibliography}
\end{document}